\pdfoutput=1
\documentclass[conference]{IEEEtran}
\usepackage{amsmath,amssymb,amsfonts}
\usepackage{booktabs}
\usepackage{graphicx}
\usepackage{bm}
\usepackage[hyphens]{url}
\begin{document}

\title{Federated Bayesian Surveillance of Mechanical Thrombectomy Adverse Events:\\A Population Risk Layer for Surgical Digital Twins}

\author{\IEEEauthorblockN{Damini Rijhwani}
\IEEEauthorblockA{Automation Core Inc., Cambridge, MA, USA\\
damini@automationcoreinc.com \quad ORCID: 0009-0000-2253-6111}}

\maketitle

\begin{abstract}
Learned surgical simulators and world models can roll out plausible procedural futures, but they carry no grounded estimate of how often interventional devices actually harm patients. We propose treating population-scale adverse-event surveillance as a distinct belief layer of the surgical digital twin, and we evaluate a federated Bayesian protocol for learning it under formal privacy guarantees. Each site holds per-class Gamma-Poisson posteriors over adverse-event rates and exchanges only R\'enyi-differentially-private natural-parameter updates. We benchmark on the complete FDA MAUDE cohort for thrombus-retrieval catheters (product code NRY): 8{,}617 reports, of which 6{,}491 are classified by transparent keyword rules into five thrombectomy complication classes and partitioned across $K{=}8$ manufacturer sites. At a matched privacy budget of $(\varepsilon{\approx}2.09, \delta{=}10^{-5})$, the conjugate protocol attains a held-out Poisson score of $-5.78$ per test event versus $-26.58$ for FedAvg with differential privacy. The non-private federated model also outperforms centralized pooling ($+3.19$ vs $+2.93$), evidence that manufacturer-specific complication profiles are real and that federation preserves them. Because MAUDE lacks procedure denominators, outputs are relative rate orderings rather than absolute risks, and we report all privacy-utility operating points.
\end{abstract}


\section{Introduction}

Surgical robotics is increasingly adopting world models, using learned simulators to predict action outcomes before execution. World foundation models now evaluate surgical robot policies in silico~\cite{zbinden2025cosmos}, digital twins are proposed to ground foundation models in operating rooms~\cite{shen2025position,ding2024robust}, and surgical digital twin research is developing rapidly~\cite{khan2025survey}. Endovascular intervention, especially mechanical thrombectomy, is a primary target of this effort.

These systems share a clear blind spot. A learned simulator can predict that a retriever will deform a vessel, but cannot estimate how often retrievers of that class perforate vessels in clinical practice, since that information resides in fragmented, sensitive post-market records rather than surgical video. Real-world failure statistics for thrombectomy devices exist in scattered form: registry studies of the FDA Manufacturer and User Facility Device Experience (MAUDE) database document device-specific complication patterns in peripheral and pulmonary procedures~\cite{ahmad2024maude,warren2026maude}, with reproducibility remaining an active concern~\cite{li2024maude}. Individual hospitals rarely see enough events to estimate rates reliably, and procedural records cannot be centralized.

We propose that surgical digital twins require a distinct \emph{population risk layer}: calibrated beliefs over adverse-event rates by device class and complication type, learned across institutions without transferring raw records, to condition per-procedure simulations. This paper introduces this layer and evaluates it on real public data:

\begin{enumerate}
\item \textbf{A federated Bayesian surveillance protocol.} Sites maintain per-class Gamma-Poisson posteriors and exchange only R\'enyi-differentially-private (RDP) natural-parameter updates of dimension $2Z$ per round. Conjugacy reduces each update to a two-scalar sufficient statistic per class instead of a gradient.
\item \textbf{A thrombectomy benchmark from complete public data.} We provide the recipe for a manufacturer-federated partition of the complete MAUDE cohort for thrombus-retrieval catheters: 8{,}617 reports, with 6{,}491 classified into five complication classes across major thrombectomy manufacturers.
\item \textbf{Evidence that federation preserves manufacturer heterogeneity.} The federated per-site model outperforms centralized pooling even in the absence of privacy noise, indicating that complication profiles differ across device makers, a structure that pooled models erase.
\end{enumerate}

\textbf{Scope.} This paper evaluates population-level rate estimation on public surveillance data. It does not evaluate per-procedure simulation, device control, or clinical decision-making, and its outputs are not evidence of comparative device safety.

\section{Federated Bayesian Risk Protocol}

\subsection{Rate model}
For complication class $z \in \{1,\dots,Z\}$ at site $k$, the event count over exposure $n_{z,k}$ is modeled as conditionally Poisson with latent rate $\theta_z$:
\begin{equation}
y_{z,k} \mid \theta_z \sim \mathrm{Poi}(n_{z,k}\,\theta_z), \qquad \theta_z \sim \mathrm{Gam}(\alpha_{z,0}, \beta_{z,0}).
\label{eq:model}
\end{equation}
The posterior update is conjugate: with natural parameter $\bm{\eta}_z = (\alpha_z - 1, -\beta_z)^{\!\top}$, a site's contribution is the sufficient statistic $\Delta\bm{\eta}_{z,k} = (y_{z,k}, -n_{z,k})^{\!\top}$, and the global posterior is the prior plus the sum of site contributions:
\begin{equation}
\bm{\eta}_z^{\mathrm{global}} = \bm{\eta}_z^{\mathrm{prior}} + \textstyle\sum_{k=1}^{K} \Delta\bm{\eta}_{z,k}.
\label{eq:agg}
\end{equation}
This estimator family extends the Bayesian shrinkage framework of pharmacovigilance signal detection~\cite{dumouchel1999,bate1998} to a multi-site setting where no party inspects the pooled table.

\subsection{Privacy and communication}
Each round, every site clips its update to norm $C$, adds Gaussian noise $\mathcal{N}(0, \sigma^2 I / K)$, and transmits under pairwise-masking secure aggregation~\cite{bonawitz2017}, ensuring the server observes only the noised sum. Sensitivity is bounded by the event-level neighboring relation: one report perturbs one class count while exposure windows remain fixed, so $\|\Delta\bm{\eta}\|_2 \le 1 \le C$. Privacy is tracked in RDP~\cite{mironov2017} with additive composition over $T$ rounds and conversion to $(\varepsilon,\delta)$~\cite{dwork2014}; sufficient-statistic perturbation for conjugate models follows~\cite{heikkila2017}. Per-round communication requires $2Z$ scalars per site, independent of cohort size, unlike gradient-based federated methods~\cite{mcmahan2017,abadi2016,kairouz2021}.

In our experiments, each round releases the statistic of an independent Bernoulli-subsampled batch (fraction $1/20$) over $T{=}200$ rounds. Counts and exposures accumulate together, so posterior means remain unbiased, and the evaluation below depends only on posterior means. Accumulated precision would overstate effective sample size by $T/20$ and must be renormalized before interpreting posterior widths. The accountant charges every release and claims no subsampling amplification.

\section{The MAUDE Thrombectomy Benchmark}

\subsection{Cohort and classification}
We retrieved the complete openFDA MAUDE cohort for product code NRY (catheter, thrombus retriever): 8{,}617 reports through August 2026, queried by authoritative product code with full pagination and archived query manifests. Report narratives are categorized by transparent, severity-ordered keyword rules into $Z{=}5$ classes: vessel perforation or dissection (1{,}293), hemorrhage (1{,}706), distal embolization (307), device fracture or detachment (2{,}806), and failure to retrieve or track (379), with 6{,}491 reports matching at least one class. Sites represent the top $K{-}1{=}7$ manufacturers by classified report volume, capturing the major thrombectomy vendors (Penumbra, Micro Therapeutics/ev3, Neuravi, Concentric Medical, Covidien, Stryker Neurovascular), plus an OTHER site ($K{=}8$). Exposure $n_{z,k}$ is the site reporting-year span, and 20\% of events per site are held out.

\textbf{Estimand.} MAUDE is voluntary surveillance lacking procedure denominators. Fitted rates represent events per reporting-year; outputs are relative rate orderings across classes and sites rather than absolute complication probabilities, and reporting propensities vary across manufacturers.

\subsection{Baselines and metric}
We compare against centralized pooled Gamma estimation, FedAvg, FedAvg with DP-SGD-style noise, FedProx, SCAFFOLD, and FedPA, all optimizing the same intercept-only Poisson objective, with $T{=}200$, $C{=}\sqrt{2}$, $\delta{=}10^{-5}$, and $\sigma \in \{6,12,24,48,96\}$. The metric is the unnormalized held-out Poisson log-likelihood per test event (higher is better); only differences across methods are interpretable. Results are reported as means over 10 seeds with 95\% bootstrap confidence intervals.

\begin{table}[t]
\centering
\caption{Held-out Poisson score per test event on the MAUDE thrombectomy federation ($K{=}8$, $Z{=}5$, 10 seeds, 95\% bootstrap CI). Higher is better; only differences are interpretable.}
\label{tab:results}
\small
\setlength{\tabcolsep}{3.5pt}
\begin{tabular}{@{}llc@{}}
\toprule
Method & Privacy & Score \\
\midrule
Centralized (pooled Gamma) & none & $+2.93\,[+2.91, +2.94]$ \\
FedAvg & none & $-26.58\,[-26.58, -26.58]$ \\
FedProx & none & $-26.58\,[-26.58, -26.58]$ \\
SCAFFOLD & none & $-26.58\,[-26.58, -26.58]$ \\
FedPA & none & $-26.58\,[-26.58, -26.58]$ \\
FedAvg$+$DP & $\varepsilon{\approx}2.09$ & $-26.58\,[-26.58, -26.58]$ \\
\midrule
Federated conjugate, non-private & none & $+3.19\,[+3.17, +3.21]$ \\
Federated conjugate, $\sigma{=}6$ & $\varepsilon{\approx}21.55$ & $+0.47\,[-0.14, +1.08]$ \\
Federated conjugate, $\sigma{=}12$ & $\varepsilon{\approx}9.39$ & $-1.16\,[-1.64, -0.69]$ \\
Federated conjugate, $\sigma{=}24$ & $\varepsilon{\approx}4.35$ & $-2.69\,[-3.31, -2.02]$ \\
Federated conjugate, $\sigma{=}48$ & $\varepsilon{\approx}2.09$ & $-5.78\,[-6.72, -4.91]$ \\
Federated conjugate, $\sigma{=}96$ & $\varepsilon{\approx}1.02$ & $-8.43\,[-10.26, -6.68]$ \\
\bottomrule
\end{tabular}
\end{table}

\begin{figure}[t]
\centering
\includegraphics[width=\columnwidth]{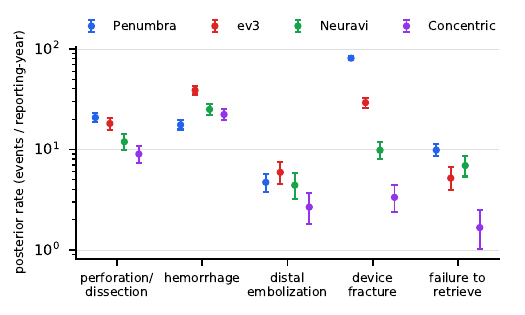}
\caption{Per-class Gamma posterior rate intervals (95\%) for the four largest manufacturer sites, non-private fit, flat prior. Rates are events per reporting-year from voluntary reports: relative orderings under differing reporting propensities, not evidence of comparative device safety. Profiles differ markedly across manufacturers, which is the heterogeneity that pooling erases.}
\label{fig:hetero}
\end{figure}

\section{Results and Discussion}

\textbf{Conjugate updates outperform gradient federation at matched privacy.} At $(\varepsilon{\approx}2.09, \delta{=}10^{-5})$, the conjugate protocol scores $-5.78$ versus $-26.58$ for FedAvg$+$DP (Table~\ref{tab:results}). Gradient baselines fail to move from their initialization on this objective; all five gradient rows, private and non-private alike, remain at $-26.58$ with degenerate confidence intervals. Because no learning-rate search was performed, these rows represent an unoptimized floor rather than tuned competitors. The supported claim is that closed-form sufficient-statistic exchange outperforms gradient optimization for this model class on these data, not that it surpasses well-tuned federated systems.

\textbf{Manufacturer heterogeneity is genuine, and federation preserves it.} The non-private federated model outperforms centralized pooling ($+3.19$ vs $+2.93$), and Fig.~\ref{fig:hetero} illustrates why. Pooling assumes a single complication profile for all devices; the data contradict this, aligning with the clinical distinction between aspiration and stent-retriever mechanisms and with device-specific MAUDE findings~\cite{ahmad2024maude,warren2026maude}. This inverts the typical framing: federation here is not a privacy-constrained compromise of a centralized ideal, but the better-specified model.

\textbf{Privacy-utility frontier.} Utility decreases monotonically with added noise, from $+0.47$ at $\varepsilon{\approx}21.55$ to $-8.43$ at $\varepsilon{\approx}1.02$. We report all operating points so deployments can select target budgets explicitly.

\textbf{Limitations.} MAUDE is voluntary and under-reports true incidence; reporting propensities vary by manufacturer and over time, and MAUDE query reproducibility is a recognized challenge~\cite{li2024maude}. Keyword classification is transparent but coarse: 25\% of reports match no class, and narrative quality varies. The privacy guarantee operates at the event level, assuming an honest-but-curious server without collusion. The evaluation scores per-site posteriors; practical deployment would require formal data-use agreements and prospective calibration assessments at the chosen operating budget.

\section{Toward Procedural Belief-State Twins}

The population layer examined here supplies priors rather than procedural dynamics. A natural extension is coupling this layer with a per-procedure belief-state twin: posterior rate beliefs conditioned on device class initialize the risk state of an intra-operative world model, which updates from sparse observations (such as pulsed fluoroscopy) and evaluates posterior-predictive risk for candidate actions before execution. That direction, aligned with twin-grounded foundation-model control~\cite{shen2025position,zbinden2025cosmos}, remains future work; this paper provides the underlying population layer learned across institutions under formal privacy guarantees.

\smallskip
\noindent\textbf{Code and data.} The benchmark recipe, classification rules, and protocol implementation will be released upon publication; all data are public (openFDA MAUDE)~\cite{fda_maude}.

\bibliographystyle{IEEEtran}
\bibliography{refs}

\end{document}